# Assessing microbiome engraftment extent following fecal microbiota transplant with q2-fmt

Chloe Herman[1,2], Evan Bolyen[1], Anthony Simard[1], Liz Gehret[1], J. Gregory Caporaso[1,2]

1. Pathogen and Microbiome Institute, Northern Arizona University, Flagstaff, AZ, USA.
2. School of Informatics, Computing and Cyber Systems, Northern Arizona University, Flagstaff, AZ, USA

## Abstract

We present q2-fmt, a QIIME 2 plugin that provides diverse methods for assessing the extent of microbiome engraftment following fecal microbiota transplant. The methods implemented here were informed by a recent literature review on approaches for assessing FMT engraftment, and cover aspects of engraftment including Chimeric Asymmetric Community Coalescence, Donated Microbiome Indicator Features, and Temporal Stability. q2-fmt is free for all use, and detailed documentation illustrating worked examples on a real-world data set are provided in the project's documentation.

## Introduction

Methods for assessing microbial engraftment following Fecal Microbiota Transplants (FMTs) are not standardized, making it difficult for researchers to investigate if FMT was an effective treatment for their condition of interest. Quantifying engraftment extent after FMTs is pivotal to interpreting the results of an FMT study. For example, if no positive clinical outcome is achieved, is it because FMT is not a relevant treatment option for the condition, or because the microbiome didn't successfully engraft?

q2-fmt is a QIIME 2 plugin for assessing microbiome engraftment extent following FMT. Accessibility of this functionality through QIIME 2 provides an all-in-one suite for assessing FMT engraftment in a software platform that many microbiome researchers are already familiar with.

Herman et al. 2024[1] defined three criteria for assessing engraftment extent. These three criteria are *Chimeric Asymmetric Community Coalescence*, *Donated Microbiome Indicator Features*, and *Temporal Stability*. Chimeric Asymmetric Community Coalescence quantifies the extent to which a recipient's microbiome composition shifts towards a donated microbiome following FMT. Donated Microbiome Indicator Features tracks specific microbiome features that are present in the donated microbiome (e.g., specific amplicon sequence variants, or a genus of interest) in the recipient's microbiome following FMT. Temporal Stability tracks the extent to which the microbiome maintains similarity to the donated microbiome following FMT. q2-fmt provides functionality to address all three of these criteria.

# Design and Implementation

## Chimeric Asymmetric Community Coalescence

### Community Coalescence (CC) Pipeline

`qiime fmt cc` is a pipeline that generates "raincloud plots" and associated statistics tracking microbiome composition changes over time. `qiime fmt cc` operates on microbiome alpha or beta diversity metrics (QIIME 2 semantic types: `SampleData[AlphaDiversity]` or `DistanceMatrix`, respectively). If a beta diversity distance matrix is passed, the user can select whether to investigate the microbiome composition distance between the recipient and either the donor or the recipient's baseline. The cc pipeline does this by running three independent actions.

First, `qiime fmt group-timepoints` prepares diversity metrics for analysis by creating a table (QIIME 2 semantic type: `Dist1D`) that contains relevant metadata (subject identifier, timepoint identifier, donor identifier) and diversity metrics (alpha diversity values, or microbiome distances to per-subject baseline or donor samples). Second, relevant statistics are computed by the q2-stats plugin (typically Wilcoxon SRT; https://github.com/qiime2/q2-stats) is run, and a statistics table (QIIME 2 semantic type: `StatsTable[Pairwise]`) is created. Lastly, the `Dist1D` and `StatsTable` are provided to `qiime stats plot-raincould`, generating a QIIME 2 Visualization containing a raincloud plot and the associated statistics (Figure 1).

### Proportional Engraftment of Donor Features (PEDF)

Proportional Engraftment of Donor Strains (PEDS)[3] was first defined by Aggarwala et al, 2020 and reports the proportion of donor features (e.g., amplicon sequence variants) that are observed in a recipient sample. We have generalized this methodology as Proportional Engraftment of Donor Features, PEDF, to enable investigation of any type of microbiome feature (e.g., OTUs, ASV, taxa, or metabolites). `qiime fmt pedf` uses microbiome feature tables (QIIME 2 semantic type: `FeatureTable`) as input and calculates the number of donor features observed in each recipient sample divided by the total number of donor features. In QIIME 2, this metric can be visualized using a heatmap by running `qiime fmt heatmap`, or using a raincloud plot (like that shown in Figure 1) to track how PEDF changes over time following an FMT. This method enables researchers to understand the proportion of donor features that were transferred to and persisted in the recipient over time.

To assess whether PEDF is identifying actual transfer from the donor to the recipient, as opposed to simply shared microbiome features between donor and recipient samples, q2-fmt introduces a novel permutation test through the `qiime fmt pedf-permutation-test` action. `pedf-permutation-test` begins by defining "mismatched donor pairs", where recipient samples are paired with donor samples from the study that don't represent their actual donor. Mismatched donor-recipient pairs are sampled with replacement from the input data, and for each pair the samples are rarified to a user-specified even sampling depth, and PEDF is computed for the recipient sample. The number of mismatched pairs sampled is defined using the `num_resamples` parameter, which defaults to 999, and rarifying is performed after sampling such that if the same mismatched pair is drawn twice the samples will be rarified independently. Then, PEDF is computed for each actual donor-recipient pair and the number of mismatched donor pairs that achieved a PEDS score greater than or equal to the actual pair plus one is divided by `num_resamples` plus one to provide a permutation test p-value for that PEDF value for that actual pair.

## Proportion of Recipients with Donor Feature (PRDF)

We additionally define a new metric related to PEDF that we refer to as Proportion of Recipients with Donor Feature, PRDF. The `prdf` action takes the same inputs as `pedf` but quantifies which features are more successful at engrafting into recipients as the proportion of recipients that successfully engrafted a specific feature when it was present in their donor as:

$$PRDF = \frac{\text{number of recipients in which a feature engrafted}}{\text{number of recipients whose donated microbiome contained feature}}$$

`qiime fmt prdf` is similarly able to be visualized using a heatmap to track which features engraft the most over time. We think that this will ultimately enable a better understanding of whether some microbes are more amenable to FMT transfer.

## Proportion of Persistent Recipient Features (PPRF)

Proportion of Persistent Recipient Features, or PPRF, is a generalized method adapted from Aggarwala et al 2021's Proportion of Persistent Recipient Strains[3] or PPRS. PPRF captures the amount of features that are persistent in the recipient following FMT. This method was calculated by the number of unique baseline (pre-FMT) features that are still present in the recipient following FMT divided by the total number of baseline features. `qiime fmt pprf` uses a feature table and a baseline timepoint to calculate the PPRF for each recipient sample. Similar to `qiime fmt pedf` this outputs a Dist1D and can be visualized using `qiime fmt heatmap`.

# Donated microbiome Indicator Features

## ANCOMBC

`qiime composition ancombc` can be used to identify donated microbiome indicator features, or features that are significantly over-represented in donated microbiome samples relative to baseline recipient samples. This enables investigation of features that are more abundant in the donor and can serve as donated microbiome indicator features. In order to facilitate the detection of donor indicator features, we present `qiime fmt detect-donor-indicators`. This pipeline takes a feature table, a metadata column with donor sample ids (which most q2-fmt actions require) and a baseline time point. This pipeline then filters the feature table down to donors and recipient baseline time point samples and runs ANCOMBC, comparing the donors to the recipients at baseline.

Future directions for identifying donated microbiome indicator features will use `qiime composition ancombc2`, which is currently in development. ANCOMBC2 models allow for repeated measurements, making it possible to run differential abundance throughout the course of the study. This would remove the need for users to filter down to their baseline and donor samples to run the comparison and will enable a more holistic comparison of donated microbiome indicator features throughout the study.

## Temporal Stability

Temporal Stability is an important aspect of a successful engraftment. With an FMT, researchers/clinicians are trying to replace the recipient's microbiome with the donated microbiome, so it is important to see that the donated microbiome is stable in the recipient following FMT rather that simply introducing a brief disturbance. Although it is common for the microbiome to drift or become personalized again after FMT (as seen in Kang et al. 2019)[4], stability of the donated microbiome is typically seen in the weeks or months following successful FMT intervention. No obvious threshold for how long the stability lasts before personalization begins has yet been identified.

q2-fmt investigates temporal stability through its focus on longitudinal data analysis in all actions. While all actions in q2-fmt can be run with just donated microbiome samples and recipient samples pre- and post-FMT, the actions are intended to be used with many recipient samples across time. This functionality allows researchers to understand the long term impacts of FMT on recipient microbiomes, and we hope that adoption of this tool and these approaches will facilitate an improved understanding of the temporal dynamics that should be expected following a successful FMT.

# Results

q2-fmt is a comprehensive plugin providing 6 actions and 2 pipelines for fully assessing engraftment extent following FMT. q2-fmt provides the tools needed for any researcher to effectively assess engraftment extent in their study participants, and facilitates the use of the same metrics for tracking microbiome engraftment extent across FMT studies. This can help standardize methodologies for assessing engraftment extent following FMT, and provide insight into the most relevant microorganisms and timeframes in FMT studies.

# Availability and Future Directions

## Testing and Maintenance

`q2-fmt` has a wide breadth of unit tests that cover all the actions found in the plugin. These unit tests are run on every commit to the `dev` branch and on every pull request.

q2-fmt can be installed in both the QIIME 2 amplicon and metagenome distributions as a community plugin using instructions in the project's documentation. Unit tests are run weekly against the newest build of the QIIME 2 amplicon distribution and the developers are alerted to test failures. This will ensure that q2-fmt remains current with the QIIME 2 framework.

q2-fmt was developed as a part of an informatics Ph.D project at Northern Arizona University. Maintenance of this software will be performed by the Caporaso Lab at Northern Arizona University. User support for q2-fmt is provided on the QIIME 2 Forum. Worked examples of applying q2-FMT to a real-world Autologous FMT (auto-FMT) cancer microbiome study are available in the project's documentation[2].

By nature of being implemented as a QIIME 2 plugin, q2-fmt is available through a Python 3 API, a command line interface, and through Galaxy's graphical user interface. Analyses are fully reproducible, as a result of the QIIME 2 framework's integrated provenance tracking and provenance replay functionality[5].

**Project name**: q2-FMT

**Project home page**: https://q2-fmt.readthedocs.io
**Operating systems**: Linux, macOS, Windows via Windows Subsystem for Linux
**Programming language**: Python 3
**Other requirements**: QIIME 2 2024.10 or later
**License**: BSD 3-Clause
**Any restrictions to use by non-academics**: None; q2-fmt is open source and free for all use.

List of works cited

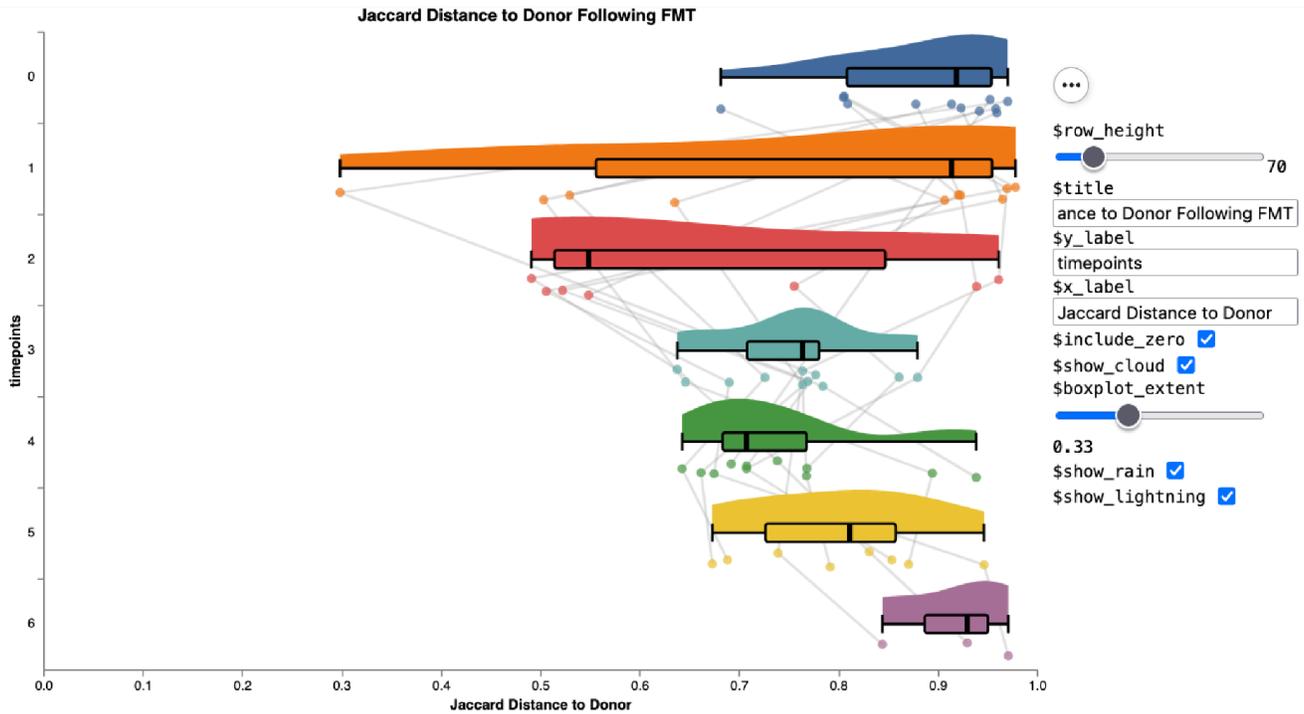

**Figure 1.** Raincloud plots showing the distribution of subjects' measure of Distance to DonorSampleID across timepoints. Kernel density estimation performed using a bandwidth calculated by Scott's method. Boxplots show the min and max of the data (whiskers) as well as the first, second (median), and third quartiles (box). Points and connecting lines represent individual subjects with a consistent jitter added across groups such that slopes across adjacent groups are visually comparable between subjects.

| Group A timepoints | Group B timepoints | A Median Distance to DonorSampleID | B Median Distance to DonorSampleID | n | test-statistic | p-value | q-value |
|---|---|---|---|---|---|---|---|
| 0 (n=12) | 1 (n=10) | 0.918839 | 0.914003 | 10 | 21 | 0.507624 | 0.507624 |
| 0 (n=12) | 2 (n=7) | 0.918839 | 0.548837 | 7 | 5 | 0.128190 | 0.192285 |
| 0 (n=12) | 3 (n=11) | 0.918839 | 0.764151 | 11 | 7 | 0.020795 | 0.062386 |
| 0 (n=12) | 4 (n=11) | 0.918839 | 0.707819 | 11 | 0 | 0.003346 | 0.020074 |
| 0 (n=12) | 5 (n=8) | 0.918839 | 0.811598 | 8 | 10 | 0.262618 | 0.315142 |
| 0 (n=12) | 6 (n=3) | 0.918839 | 0.929825 | 3 | 0 | 0.108809 | 0.217619 |

**Figure 1. Jaccard Distance to Donor following FMT intervention, as assessed using `qiime fmt cc`.** This Visualization allows for users to interactively investigate whether recipient microbiomes become more similar to donor microbiomes with treatment. In this figure, FMT intervention was at time point 3. Since this is an autoFMT study, the 'donor' is the recipient before cancer treatment. This figure shows that spontaneous recovery from cancer treatment leads to some individuals recovering quickly while others do not. The FMT intervention decreases the variability in distance to donor of all recipients. We see that their distance to donor increases at the last time point, but this is up to 2 years after the initial FMT and may indicate the development of an "individualized microbiome". Presented data is from Taur *et al.* (2018)[2]. An interactive version of this figure is available in the project documentation at https://q2-fmt.readthedocs.io.